# Quantum Conductivity for Metal-Insulator-Metal Nanostructures

Joseph W. Haus[1,2], Domenico de Ceglia[2], Maria Antonietta Vincenti[2], Michael Scalora[3]

[1] *Electro-Optics Program, University of Dayton, Dayton, OH 45469-2951*
[2] *National Research Council - AMRDEC, Charles M. Bowden Research Center, Redstone Arsenal, AL 35898*
[3] *Charles M. Bowden Research Center, AMRDEC, RDECOM, Redstone Arsenal, AL 35898-5000*
e-mail: jwhaus@udayton.edu

**ABSTRACT**

We present a methodology based on quantum mechanics for assigning quantum conductivity when an ac field is applied across a variable gap between two plasmonic nanoparticles with an insulator sandwiched between them. The quantum tunneling effect is portrayed by a set of quantum conductivity coefficients describing the linear ac conductivity responding at the frequency of the applied field and nonlinear coefficients that modulate the field amplitude at the fundamental frequency and its harmonics. The quantum conductivity, determined with no fit parameters, has both frequency and gap dependence that can be applied to determine the nonlinear quantum effects of strong applied electromagnetic fields even when the system is composed of dissimilar metal nanostructures. Our methodology compares well to results on quantum tunneling effects reported in the literature and it is simple to extend it to a number of systems with different metals and different insulators between them.

# 1. INTRODUCTION

Nanoplasmonics has emerged as a burgeoning field of research with potential applications covering broad areas of technology including electronics, medicine, environment and renewable energy. The traditional approach to our understanding of nanoplasmonic systems problems involves comprehensive electromagnetic simulations of the nanostructures using classical expressions for the dielectric response of the materials. The optical or purely electromagnetic properties of plasmonic structures for a wide range of applications have been studied in many publications [1-15]. The classical approach has worked well for many systems whose dimensions are greater than a few nanometers. However, there are proposed systems and applications where the minimum feature sizes are reduced below a nanometer. This situation has triggered greater scrutiny of the assumptions underlying the classical treatment of plasmons.

To overcome the limitations inherent in the classical nanoplasmonic analysis a sub-field of nanophotonics, sometimes referred to as quantum plasmonics, recently emerged. One branch of this sub-field explores the quantum tunneling of electrons between nanostructured metals (plasmonics) that are separated by insulator materials. The quantum techniques range from numerically intensive time-dependent density functional theory to a simple Quantum Correction Model that computes a linewidth factor or spatial dispersion models; the latter approaches are based on the Drude model [16-22]. While electron tunneling is driven by electromagnetic fields, charge exchange limits field enhancement in the gap region between the two metals. In contrast, the purely classical models predict an ever increasing local field enhancement as the gap is reduced. Recent papers in quantum plasmonics have explored the effect of electron tunneling due to metal dimers in vacuum.

Our work draws from the literature developed for metal-insulator-metal (MIM) structures [23-36]. A quantum mechanical model is applied that describes tunneling with an applied time-dependent field. Then a set of frequency-dependent quantum conductivity coefficients is extracted for the nanoplasmonic system. By applying standard quantum mechanical techniques the ac or dc currents enabled via the electron tunneling effect can be estimated by endowing the insulator volume with a set of quantum conductivity functions. We call this approach the Quantum Conductivity Theory (QCT) and it can be applied to nanoplasmonic systems composed of different metals and having different insulator materials separating the metals. Nonlinear quantum conductivity coefficients will enable harmonic generation and two-photon absorption-like phenomena. QCT can be incorporated into computational electromagnetic models in a straightforward fashion, as we demonstrate in this paper.

To illustrate how the quantum conductivity affects electromagnetic scattering in nanosystems a system of metal dimers is considered. We apply QCT to two geometries of gold dimers immersed in different media with a variable gap between the metal structures. When vacuum surrounds the dimers the results are compared to available simulation data. To establish the versatility of QCT we embed the gold dimers in insulators and the results are strikingly different. We also examine the case of a sodium dimer in vacuum for comparison with recent results and the sodium dimer system is used to demonstrate that nonlinear effects also contribute to diminishing the field enhancement. QCT modifies both the linear and nonlinear absorption of the electromagnetic fields and generates harmonic fields.

## 2. QUANTUM TUNNELING ANALYSIS

Our analysis begins by considering a free electron wave driven by an applied voltage and moving through a region with a barrier potential. The current is a combination of the electron tunneling probability and the occupation densities of the energy levels. The standard form of Schrödinger's equation is [37]

$$E\psi = -\frac{\hbar^2}{2m}\nabla^2\psi + V(x,y,z)\psi \, . \tag{1}$$

We solve Schrödinger's equation to determine the wave function using the transfer matrix method. Two contributions to the potential are considered [38, 39]. The first is related to the electronic material parameters and the second is the electrostatic force on the tunneling electrons due to the metal walls, so that one can write:

$$V(x,y,z) = V_{material}(x,y,z) + V_{image}(x,y,z) \, . \tag{2}$$

For the single-electron description the material portion of the potential function is determined by two coefficients. One parameter is the work function, $W$, which is the minimum kinetic energy the electron requires to escape from the metal into the vacuum. The other parameter is the electron affinity, $\Phi$, which measures the ability of the insulator to capture an electron at rest from the vacuum to the bottom of the conduction band. These two parameters are listed for several metals and insulators in Table 1. The values are quoted in electron volts.

Table 1: Selected material property values for metals and insulators [39].

| Metal | W [eV] | Insulator | Φ [eV] | K |
|---|---|---|---|---|
| Au | 5.1 | $TiO_2$ | 3.9 | 7.8 |
| Na | 2.75 [40] | $SiO_2$ | 0.9 | 2.25 |

The barrier height at the interface between the metal and insulator is given by

$$\varphi = W - \Phi \tag{3}$$

The potential barrier due to the materials parameters with the insulator region in the center between two different metals is illustrated in Figure 1. The barrier height is measured from the Fermi energy, which is equal in both metals for the equilibrium case. The applied voltage, $\bar{V}_d$, is indicated by lowering the Fermi level of one metal, assuming the other metal is grounded. In the following voltages are distinguished from a potential by a bar over the letter *V*.

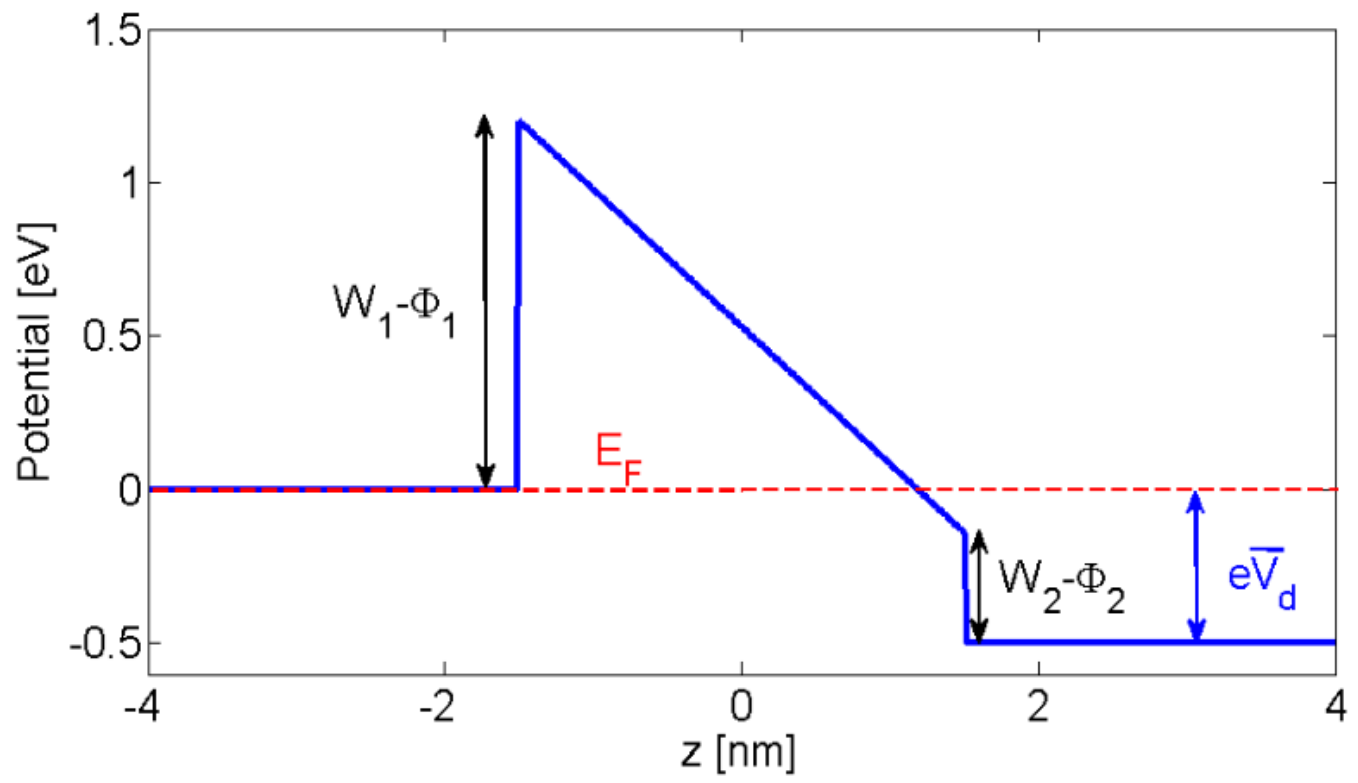

Figure 1: Illustration of the potential barrier in the MIM structure. In equilibrium the Fermi energy is the same in both metals. The application of an external voltage depresses the Fermi energy in one metal from its equilibrium value.

We apply our methodology to two metals, gold and sodium, because there are results from prior literature and we can compare our results with them [18, 22]. The electron affinities in Table 1 represent values from zero to a relatively large electron affinity value for titanium dioxide. We note that the sodium work function in Table 1 was selected from published the literature [40], but we found other reported values.

The image potential represents the interaction of the electron in the barrier region with the metal walls. The image potential problem is determined by using an infinite array of charges displaced from one another on a one-dimensional lattice [38, 39]

$$V_{image}(x,y,z) = -\frac{e^2}{4\pi K\varepsilon_0}\left[\frac{1}{2z} + \sum_{n=1}^{\infty}\left(\frac{nd}{(nd)^2 - z^2} - \frac{1}{nd}\right)\right] \approx -\frac{1.15\ln 2 e^2 d}{8\pi K\varepsilon_0 z(d-z)}, \qquad (4)$$

where $e$ is the charge of the electron, $\varepsilon_0$ is the dielectric permittivity of free space, $K$ is the relative dielectric constant of the insulator, and $d$ is the gap distance. The relative dielectric constants for two selected insulators are listed in Table 1. The last equality is an approximation that accurately captures the summation. The total potential including the image potential is plotted in Figure 2. It has the effect of rounding the sharp edges of the potential. The two solutions in Eq. (4) for the image charge are close enough to be indistinguishable in the numerical calculations of the current density. Both formulas are easy to implement in numerical calculations.

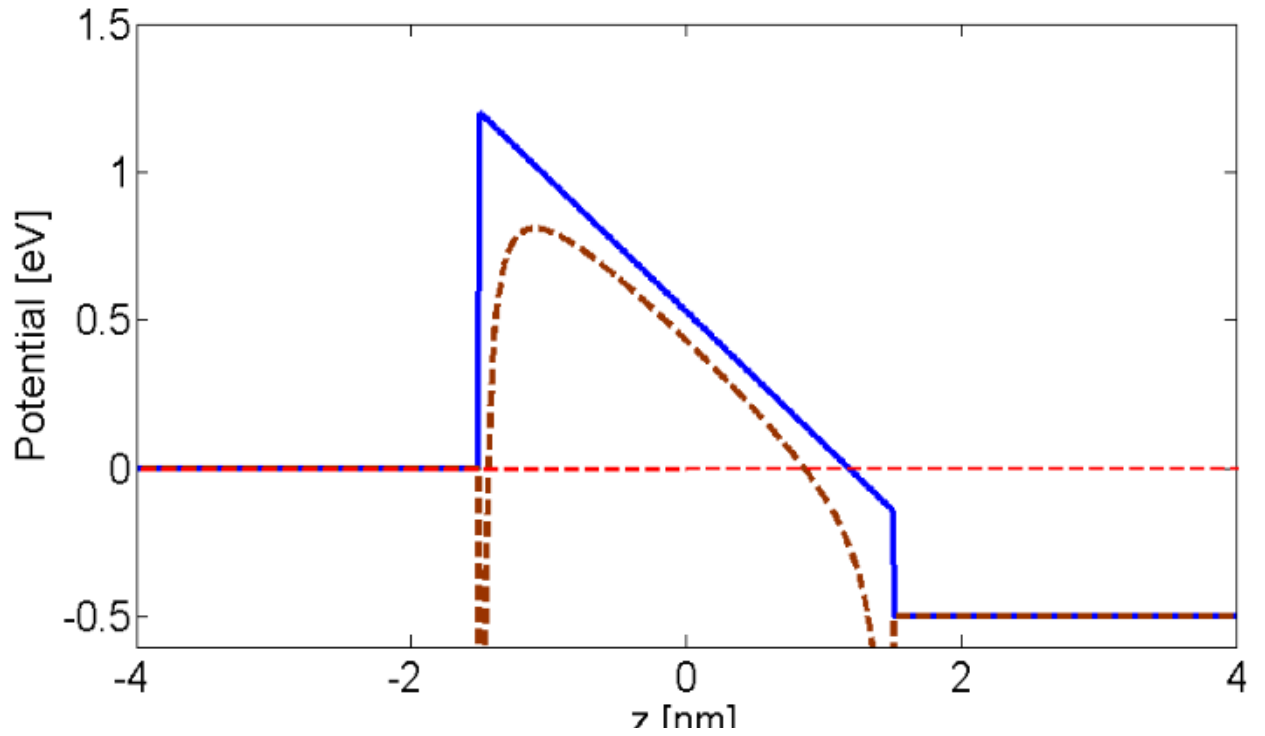

Figure 2: An illustration of two potential functions. The potential labeled with the dashed (brown) curve includes the image potential [38, 39]. The potential plotted with the solid line (blue) is the form without the image charge for comparison.

The expression for the current density [38, 39] for one direction of the applied voltage is

$$J_{dc}(e\bar{V}_d) = 4\pi e\frac{m}{h^3}\int_0^{\infty} dE_z T(E_z)\int_{E_z}^{\infty}(f(E) - f(E + e\bar{V}_d))dE\,, \qquad (5)$$

where $E_z$ is the longitudinal energy of the electron and is the total electron energy. The Fermi distribution functions are

$$f(E) = \frac{1}{e^{(E-E_F)/k_BT} + 1}, \qquad (6)$$

where $E_F$ is the Fermi energy, $T$ the lattice temperature and $k_B$ is Boltzmann's constant. We use room temperature (300 K) in all our calculations.

The Transfer Matrix Method (TMM) is implemented by slicing the potential for any shape potential into a set of segments and in each one the potential is taken as constant. Schrödinger's equation is solved by applying the boundary conditions at each interface. The result for the wave function is used to calculate the transmission function $T(E_z)$ [36].

Figure 3 is an illustration of the current flow when a voltage bias is applied for gold with three different insulators, namely vacuum, $SiO_2$ and $TiO_2$. The current map plotted in Figure 3 is the unilluminated or dark current characteristic from Eq. (5). The material electron affinities and their dielectric constants are listed in Table 1. The currents are anti-symmetric with respect to the applied voltage sign. The voltage induces a current to flow in one direction or the other by lowering the Fermi energy of one metal. The electrons tunnel through the barrier in one direction inducing a current in the opposite one. The average current depends on the quantum mechanical current weighted by the occupation of the electronic states on both side of the barrier.

We note a difference between the magnitudes of the currents spanned for the first two cases and the third one. This is because the barrier height for vacuum and $SiO_2$ gap is large compared with the one for the $TiO_2$ gap. For vacuum and $SiO_2$ the current density is suppressed until the gap becomes thin enough to produce a high tunneling probability current predominantly caused by a lowering of the barrier due to the electrostatic potential. From Table 1 we note that the larger electron affinity of $TiO_2$ lowers the barrier by 3 eV to almost 4 eV when compared with the two other cases. The barrier reduction dramatically increases the current density for large gap sizes, as seen by the range of current magnitudes in the colorbar scales to the right of each map in Figure 3; the current density for the titanium dioxide case is twenty orders of magnitude larger at the lower end of the scale. Large applied voltages are required to calculate the illuminated current density, as will be seen in the following section.

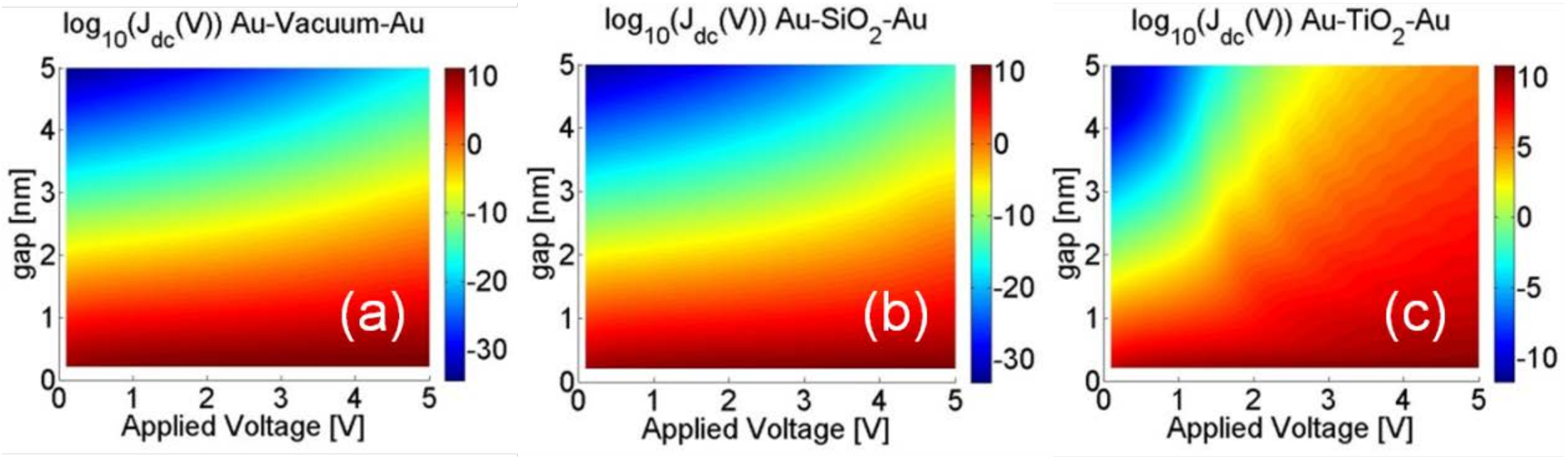

Figure 3: Maps of the logarithm (base 10) of the tunneling unilluminated current versus applied voltage and gap thickness for three cases: (a) Au/vacuum/Au, (b) Au/$SiO_2$/Au and (c) Au/$TiO_2$/Au (Right). The minimum gap is 0.2 nm.

## 3. TIME-DEPENDENT CURRENT

Tien and Gordon derived results for a photon-assisted theory of tunneling applied to normal or superconductor MIM structures [42]. It was based on the observation that the wave function for the electrons in the metal on one side of the tunneling barrier may be written as:

$$\psi(t) = e^{-i\int V(t')dt'/\hbar}\psi_0(t), \tag{7}$$

where $\psi_0(t)$ is the solution without the applied electromagnetic field and the field has a dc and harmonic component: $V(t) = e\overline{V}_d + \left(\frac{e\overline{V}_\omega}{2}e^{-i\omega t} + c.c.\right)$ and the complex amplitude is expressed as: $\overline{V}_\omega = |\overline{V}_\omega| e^{i\phi}$. The propagator is defined as

$$U(t) = e^{-i\int V(t')dt'/\hbar}, \tag{8}$$

and it can be written as a product of two terms with the second one being periodic in time

$$U(t) = U_0 U_P = e^{-ie\overline{V}_d t/\hbar - (\overline{V}_\omega - \overline{V}_\omega^*)/2\hbar\omega} e^{-(\overline{V}_\omega e^{-i\omega t} - \overline{V}_\omega^* e^{i\omega t})/2\hbar\omega}. \tag{9}$$

The overall phase factor in Eq. (9) can be neglected. Tucker wrote a complete description of the theory and here the elements are written to identify the complex field components of the results [43-45]. The current density using the equilibrium Green function theory can be expressed as:

$$J(t) = \mathrm{Im}\left\{\int\int d\omega' d\omega'' W(\omega')W(\omega'')e^{-i(\omega'-\omega'')t} j(\hbar\omega' + e\overline{V}_d)\right\}, \tag{10}$$

where $j(\hbar\omega' + e\overline{V}_d)$ is the complex current density related to the dc current density by the relationship:

$$J_{dc}(\hbar\omega' + e\overline{V}_d) = \mathrm{Im}\left\{j(\hbar\omega' + e\overline{V}_d)\right\}. \tag{11}$$

The spectral density function, $W(\omega')$, is related to the propagator by a Fourier transform

$$U_p(t) = \int_{-\infty}^{\infty} d\omega' W(\omega')e^{-i\omega' t}, \tag{12}$$

which for the periodic propagator in Eq. (10) is given by the expression

$$W(\omega') = \sum_{n=-\infty}^{\infty} J_n(\alpha)e^{in\phi}\delta(\omega' - n\omega). \tag{13}$$

$J_n(\alpha)$ is the Bessel function of order n. The variable $\alpha = e\overline{V}_\omega / \hbar\omega$ appearing in its argument is the ratio of the ac voltage to the photon voltage ($\overline{V}_{photon} = \hbar\omega / e$). $V_\omega$ is the ac voltage across the insulator gap of the MIM structure. The potential $\overline{V}_\omega$ is related to the strength of the local electric field, $E_\omega$, by the relation

$$\overline{V}_\omega = E_\omega d \ , \tag{14}$$

where *d* is the thickness of the insulating gap. If the electric field amplitude varies along the path then the voltage may be evaluated as the line integral of the field along the path of the field line. Typically, the variable $\alpha$ is small and an expansion can be made. The result for the time-dependent current density in the form of a Fourier series is:

$$J(t) = \sum_{m=0}^{\infty}\left(\frac{J_{m\omega}}{2}e^{-im\omega t} + c.c.\right) = \mathrm{Re}\left\{\int\int d\omega' d\omega'' W(\omega')W(\omega'')e^{-i(\omega'-\omega'')t}J_{dc}(\hbar\omega' + e\overline{V}_d)\right\}. \tag{15}$$

The rectified contribution (m=0) to the illuminated tunneling current density is given by

$$J_{rect} = \sum_{n=-\infty}^{\infty} J_n^2(\alpha) J_{dc}(n\hbar\omega + e\overline{V}_d)\,, \tag{16}$$

and the complex, frequency-dependent current amplitudes are expressed in general as

$$J_{m\omega} = \sum_{n=-\infty}^{\infty} J_n(\alpha)[J_{n+m}(\alpha) + J_{n-m}(\alpha)]e^{im\phi} J_{dc}(n\hbar\omega + e\overline{V}_d)\,. \tag{17}$$

For definiteness we write the lowest order contributions for the coefficients of the rectified current. The Bessel functions are expanded to retain all terms up to second order in $\alpha$. The time-averaged current term can be expanded to second order in the applied field by taking three terms in the series *n* = 0, 1 and -1,

$$\begin{aligned} J_{rect} &= J_0{}^2(\alpha)J_{dc}(e\overline{V}_d) + J_1{}^2(\alpha)\left(J_{dc}(\hbar\omega + e\overline{V}_d) + J_{dc}(-\hbar\omega + e\overline{V}_d)\right) \\ &\approx \left(1 - \frac{\alpha^2}{2}\right)J_{dc}(e\overline{V}_d) + \frac{\alpha^2}{4}\left(J_{dc}(\hbar\omega + e\overline{V}_d) + J_{dc}(-\hbar\omega + e\overline{V}_d)\right) \end{aligned} \,. \tag{18}$$

The nonlinear rectified current contribution vanishes when no bias voltage is applied and the two metals are identical. Otherwise, the nonlinear response is attributable to a rectification of the illuminated field, which has been applied to calculate the responsivity and energy harvesting efficiency [33-36] of optical rectennas. For the dissipative, in-phase response of the ac current at the fundamental frequency the leading terms to third-order in the applied field are

$$
\begin{aligned}
J_{\omega} = &\left\{ J_1(\alpha)\left(J_2(\alpha)+J_0(\alpha)\right)\left( J_{dc}(\omega+\frac{e\overline{V}_d}{\hbar}) - J_{dc}(-\omega+\frac{e\overline{V}_d}{\hbar}) \right) + \right.\\
&\left. + J_1(\alpha)J_2(\alpha)\left( J_{dc}(2\omega+\frac{e\overline{V}_d}{\hbar}) - J_{dc}(-2\omega+\frac{e\overline{V}_d}{\hbar}) \right) \right\} e^{i\varphi}\\
&\approx \left\{ \frac{\alpha}{2}\left(1-\frac{\alpha^2}{8}\right)\left(1+\mathrm{O}(\alpha)^4\right)\left( J_{dc}(\omega+\frac{e\overline{V}_d}{\hbar}) - J_{dc}(-\omega+\frac{e\overline{V}_d}{\hbar}) \right) \right.\\
&\left. + \frac{\alpha}{2}\left(\frac{\alpha^2}{4}\right)\left( J_{dc}(2\omega+\frac{e\overline{V}_d}{\hbar}) - J_{dc}(-2\omega+\frac{e\overline{V}_d}{\hbar}) \right) \right\} e^{i\varphi}\\
&\approx \frac{\alpha e^{i\varphi}}{2}\left( J_{dc}(\omega+\frac{e\overline{V}_d}{\hbar}) - J_{dc}(-\omega+\frac{e\overline{V}_d}{\hbar}) \right)\\
&+ \frac{\alpha e^{i\varphi}}{2}\left(\frac{\alpha^2}{8}\right)\left( 2J_{dc}(2\omega+\frac{e\overline{V}_d}{\hbar}) - 2J_{dc}(-2\omega+\frac{e\overline{V}_d}{\hbar}) - J_{dc}(\omega+\frac{e\overline{V}_d}{\hbar}) + J_{dc}(-\omega+\frac{e\overline{V}_d}{\hbar}) \right).
\end{aligned}
\tag{19}
$$

The first term is linear and it is related to the dissipative power of the MIM diode, but there is a third-order nonlinear dissipative contribution to the current that modulates the current amplitude. The next terms in the series (m=2,3,…) in Eq. (17) are harmonic contributions to the current; the second-harmonic (m=2) and third-harmonic (m=3) contributions are nonlinear and their lowest-order approximations are

$$
\begin{aligned}
J_{2\omega} = &\left\{ 2J_0(\alpha)J_2(\alpha)J_{dc}(\hbar e\overline{V}_d) \right.\\
&+J_1(\alpha)\left(J_3(\alpha)-J_1(\alpha)\right)\left(J_{dc}(\hbar\omega+e\overline{V}_d)+J_{dc}(-\hbar\omega+e\overline{V}_d)\right)\\
&\left. +J_2(\alpha)\left(J_4(\alpha)+J_0(\alpha)\right)\left(J_{dc}(2\hbar\omega+e\overline{V}_d)+J_{dc}(-2\hbar\omega+e\overline{V}_d)\right) \right\} e^{i2\varphi}\\
&\approx \left\{ \frac{\alpha^2}{4}J_{dc}(e\overline{V}_d) - \frac{\alpha^2}{4}\left(J_{dc}(\hbar\omega+e\overline{V}_d)+J_{dc}(-\hbar\omega+e\overline{V}_d)\right) \right.\\
&\left. +\frac{\alpha^2}{8}\left(J_{dc}(2\hbar\omega+e\overline{V}_d)+J_{dc}(-2\hbar\omega+e\overline{V}_d)\right) \right\} e^{i2\varphi}.
\end{aligned}
\tag{20}
$$

and

$$
\begin{aligned}
J_{3\omega} &= \Big\{ J_1(\alpha)\big(J_4(\alpha)+J_2(\alpha)\big)\big(J_{dc}(\hbar\omega+e\overline{V}_d)-J_{dc}(-\hbar\omega+e\overline{V}_d)\big) \\
&\quad + J_2(\alpha)\big(J_5(\alpha)-J_1(\alpha)\big)\big(J_{dc}(2\hbar\omega+e\overline{V}_d)-J_{dc}(-2\hbar\omega+e\overline{V}_d)\big) \\
&\quad + J_3(\alpha)\big(J_6(\alpha)+J_0(\alpha)\big)\big(J_{dc}(3\hbar\omega+e\overline{V}_d)-J_{dc}(-3\hbar\omega+e\overline{V}_d)\big)\Big\} e^{i3\varphi} \\
&\approx \Big\{ \frac{\alpha}{2}\frac{\alpha^2}{8}\big(J_{dc}(\hbar\omega+e\overline{V}_d)-J_{dc}(-\hbar\omega+e\overline{V}_d)\big) \\
&\quad - \frac{\alpha^2}{8}\frac{\alpha}{2}\big(J_{dc}(2\hbar\omega+e\overline{V}_d)-J_{dc}(-2\hbar\omega+e\overline{V}_d)\big) \\
&\quad + \frac{\alpha^3}{48}\big(J_{dc}(3\hbar\omega+e\overline{V}_d)-J_{dc}(-3\hbar\omega+e\overline{V}_d)\big)\Big\} e^{i3\varphi}.
\end{aligned}
\tag{21}
$$

The second-harmonic term vanishes when the bias voltage is zero in a MIM structure made with the same metals. The last term in Eq. (21) would normally be the strongest contribution especially in cases where the photon voltage corresponds to a significant fraction of the applied voltage. The third-harmonic wave is generated even in a MIM made from the same metals and without a bias voltage applied. The asymmetry made by applying a dc field or by using metals with different work functions will lead to a manifestation of nonlinear effects in these and other harmonics. As mentioned above the parameter $\alpha$ is generally small for these systems even when a nanoscale gap is used between the two metals.

## 4. QUANTUM CONDUCTIVITY COEFFICIENTS

The current flow induced by an applied field forms the basis for calculating the conductivity of a medium. As we saw in the prior section the current induced from a time varying electromagnetic field is determined from the unilluminated current in Eq. (5). The relation between the current density and the field can be used to identify nonlinear conductivities in the medium. We define the following coefficients:

$$
\begin{aligned}
J_{rect} &= J_{dc}(e\overline{V}_d)+\sigma_0^{(2)}\left|E_\omega\right|^2, \\
J_\omega &= \sigma_\omega^{(1)}E_\omega+\sigma_\omega^{(3)}\left|E_\omega\right|^2 E_\omega, \\
J_{2\omega} &= \sigma_{2\omega}E_\omega{}^2, \\
J_{3\omega} &= \sigma_{3\omega}E_\omega{}^3.
\end{aligned}
\tag{22}
$$

The five coefficients are directly extracted from Eqs. (18-21):

$$\sigma_0^{(2)} = \left(\frac{ed}{2\hbar\omega}\right)^2 \left(J_{dc}(\hbar\omega + e\overline{V}_d) + J_{dc}(-\hbar\omega + e\overline{V}_d) - 2J_{dc}(e\overline{V}_d)\right),$$

$$\sigma_\omega^{(1)} = \left(\frac{ed}{2\hbar\omega}\right)\left(J_{dc}(\hbar\omega + e\overline{V}_d) - J_{dc}(-\hbar\omega + e\overline{V}_d)\right),$$

$$\sigma_\omega^{(3)} = \left(\frac{ed}{2\hbar\omega}\right)^3 \left(J_{dc}(2\hbar\omega + e\overline{V}_d) - J_{dc}(-2\hbar\omega + e\overline{V}_d) - J_{dc}(\hbar\omega + e\overline{V}_d) + J_{dc}(-\hbar\omega + e\overline{V}_d)\right),$$

$$\sigma_{2\omega} = \left(\frac{ed}{2\hbar\omega}\right)^2 \left(\frac{1}{2} J_{dc}(e\overline{V}_d) - \left(J_{dc}(\hbar\omega + e\overline{V}_d) + J_{dc}(-\hbar\omega + e\overline{V}_d)\right)\right. \tag{23}$$
$$\left. + \frac{1}{2}\left(J_{dc}(2\hbar\omega + e\overline{V}_d) + J_{dc}(-2\hbar\omega + e\overline{V}_d)\right)\right),$$

$$\sigma_{3\omega} = \frac{1}{2}\left(\frac{ed}{2\hbar\omega}\right)^3 \left(\left(J_{dc}(\hbar\omega + e\overline{V}_d) - J_{dc}(-\hbar\omega + e\overline{V}_d)\right) - \left(J_{dc}(2\hbar\omega + e\overline{V}_d) - J_{dc}(-2\hbar\omega + e\overline{V}_d)\right)\right.$$
$$\left. + \frac{1}{3}\left(J_{dc}(3\hbar\omega + e\overline{V}_d) - J_{dc}(-3\hbar\omega + e\overline{V}_d)\right)\right)$$

The leading term in brackets has units of inverse electric field strength; the appropriate power of this quantity sets a scale for the nonlinear coefficients. For instance, a wavelength of 1 micron and a gap of 1 nm yield the field strength $\frac{\hbar\omega}{ed} = 1.24 \cdot 10^9$ V/m. This is indeed a large value when compared even against electric field values from intense pulsed laser beams. The linear ac conductivity, $\sigma_\omega^{(1)}$, constitutes a resistive response to the applied electromagnetic field. The nonlinear terms are small, but they can have a strong effect on the field strength in the gap region. There are four nonlinear conductivities in Eqs. (23), but higher order terms can be derived by extension of the field expansion.

The rectification term and the second-harmonic term are both quadratic in the field. The rectification term is used throughout the literature to calculate the change in a dc current under illumination. The term $\sigma_\omega^{(3)}$ has the effect of a two-photon absorption (TPA) contribution in Maxwell's equations. It is normally positive, which increases the absorption of light. The last two quantum conductivities describe second-harmonic and third-harmonic generation. Higher-order harmonic terms can be derived by extending the expansion and the additional conductivities could be utilized to study high-harmonic generation in nanosystems.

## 5. RESULTS

### *5.1 Quantum Conductivities Examples*

The current maps in Figure 3 bear out that the current saturates as the gap size vanishes. When the space separating the metals is a vacuum the potential barrier is highest and so the range of current density values also spans many decades over the span of gap sizes. Titanium dioxide has a high electron affinity and the range of current values for the same span of gaps is greatly reduced; furthermore, the dependence of the current on the gap distance is much weaker.

Similarly, the linear ac conductivity coefficients for gold electrodes with three different insulating layers sandwiched between them have maps that vary over a large range of values (see Figure 4); however, very small values of the conductivities have no discernible effect on the electromagnetic field. It is not until the distance is on the order of a nanometer that the quantum properties become important enough to measurably affect the electromagnetic field; we find that these values of the quantum conductivity are of order 0.1 S/m. The gap distance for such values is around 1 nm for the cases with vacuum and silica, but it extends to larger gap sizes in the titanium dioxide case, as expected from the previous discussion of barrier heights. The range of conductivities in Figure 4, as seen in the colorbars on the right of each map, is much smaller when the insulator is titanium dioxide. As mentioned previously, this is due to the large electron affinity of titanium dioxide.

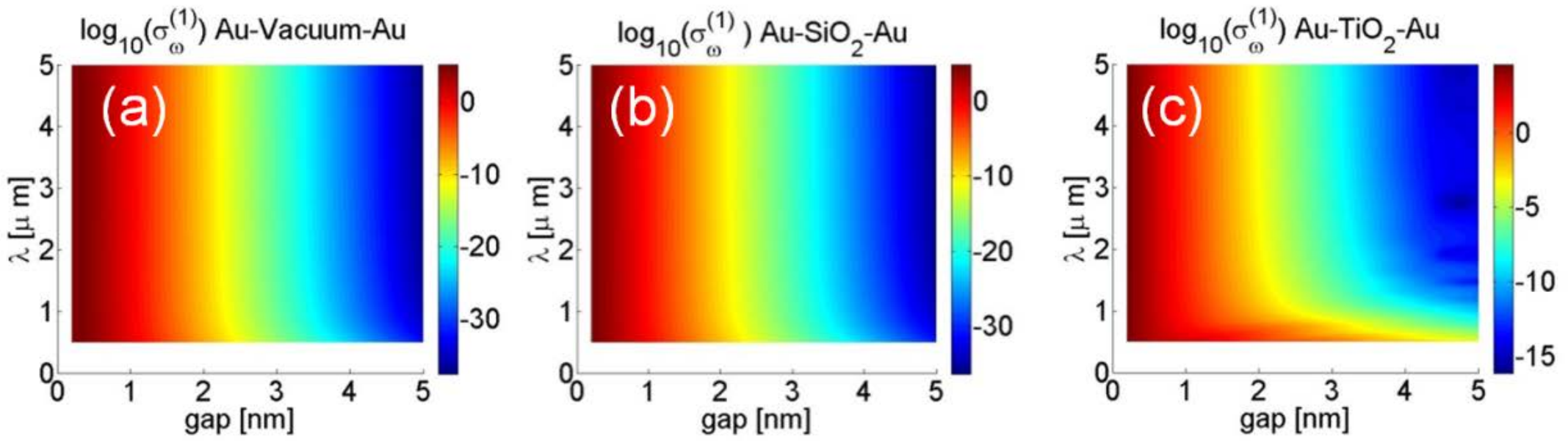

Figure 4: Maps of the logarithm (base 10) of the linear ac conductivity, $\sigma_{\omega}^{(1)}$, versus applied voltage and gap thickness for three cases: (a) Au/vacuum/Au, (b) Au/$SiO_2$/Au and (c) Au/$TiO_2$/Au. The units of the linear conductivity are S/m.

There are two non-vanishing nonlinear conductivity coefficients in MIM systems made from the same metals. They are the TPA conductivity and the third harmonic generation (THG) conductivity coefficients, which are plotted in Figures 5 and 6. As for the linear ac conductivity coefficient, the maps of the nonlinear conductivities demonstrate the same trend for different insulators. The vacuum and silica insulator cases are similar in both magnitude and wavelength dispersion. Over the range of wavelengths used in the plots, the conductivities show a relatively weak dependence on wavelength across the visible to mid-wave infrared regions when the gap size is small. The results with the dimer immersed in titanium dioxide are quite distinct from the other two host insulators; the minimum values of the conductivities are twenty orders of magnitude larger in $TiO_2$, but the maxima are smaller by a few orders of magnitude. Most of the differences are directly attributable to the difference in the electron affinity for the three cases, although the dielectric constants play a role in the observed trends too.

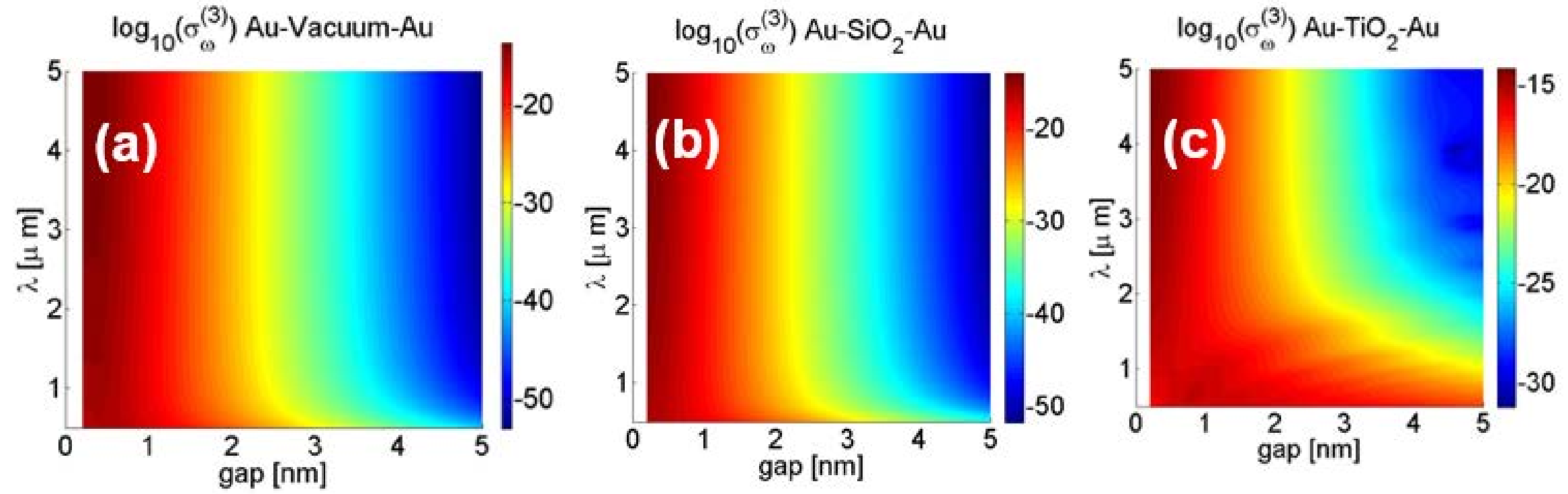

Figure 5: Maps of the logarithm (base 10) of the nonlinear TPA conductivity, $\sigma_{\omega}^{(3)}$, versus applied voltage and gap thickness for three cases: (a) Au/vacuum/Au, (b) Au/$SiO_2$/Au and (c) Au/$TiO_2$/Au. The units of the nonlinear conductivity are S m/$V^2$.

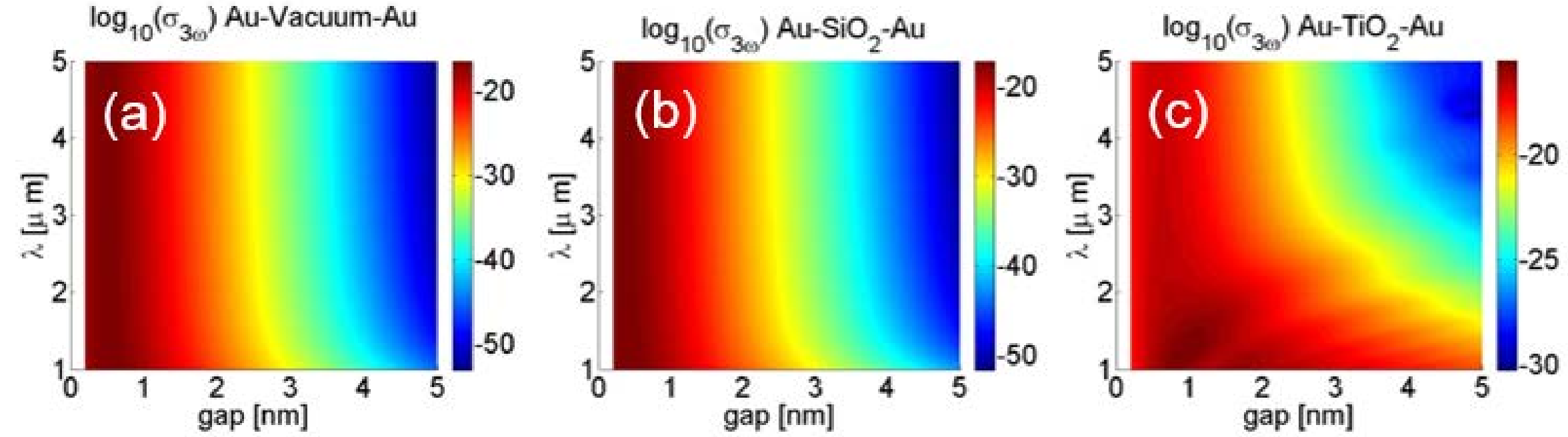

Figure 6: Maps of the logarithm (base 10) of the third-harmonic conductivity, $\sigma_{3\omega}$, versus applied voltage and gap thickness for three cases: (a) Au/vacuum/Au, (b) Au/$SiO_2$/Au and (c) Au/$TiO_2$/Au. The units of the nonlinear conductivity are S m/$V^2$.

### *5.2 Application to Gold Dimers*

The application of the quantum conductivities requires endowing the space between the metals with its physical properties. The following simple prescription can be adopted from different directions for assigning the quantum conductivities across the insulator. It is convenient and simple to implement, but not the only way. The gap between two metallic nanostructures is the key parameter for determining the distribution of conductivities. The distance from one point to another on two neighboring metal surfaces can be adopted as an estimate of the quantum conductivities along that line. We assume the following assignment for simple dimers. Lines can be extended from one metal to another along the symmetry axis; they form annulus-shaped shells in which with quantum conductivities in each shell is determined by the gap distance from one metallic boundary to the other. The width of the annulus is small compared to changes in the quantum conductivity with gap distance. When the gap becomes large enough the quantum conductivities become so small that electronic tunneling is negligible.

Using this procedure several applications of the QCT to nanosystems are considered and we compare our results to relevant literature where possible. The first example is a three-dimensional system of gold cylinder dimers. The total length of the gold cylinder is 100 nm and its radius is 5 nm. There is a gap cut out of its center and the gap is varied between 0.2 nm and 5 nm; the gap is vacuum filled. A plane electromagnetic wave with the electric field parallel to the cylinders' axis impinges on the antenna as shown in Figure 7(a). The quantum conductivity for gold-vacuum-gold dimer is applied in the gap region and the conductivity is zero outside the gap. The maximum field enhancement in the center of the gap region is evaluated by using the circuit-theory-based procedure outlined in Ref. [46]. Briefly, the intrinsic dipole impedance, $Z_{dip}$, is first retrieved by exciting the antenna with a delta-gap source [47], i.e., by establishing a constant voltage across the gap and measuring the induced current at the source position. This evaluation is performed with a 2D finite element solver (COMSOL Multiphysics) by taking advantage of the axial symmetry of the cylindrical antenna.

The gap impedance can be straightforwardly written as $Z_{gap} = -id / \left( \omega \varepsilon_0 \varepsilon_{gap}^{*} \pi r^2 \right)$, where $\varepsilon_{gap}^{*} = \varepsilon_{gap} - i\sigma_{\omega}^{(1)} / \left( \omega \varepsilon_0 \right)$ is the complex, relative permittivity in the gap region in the quantum regime, $\varepsilon_{gap}$ is the gap dielectric constant in the classical limit and $\sigma_{\omega}^{(1)}$ is defined in Eq. (23) is also constant in the gap region and taken as zero outside the gap. The electric field in the gap region is given by $E_{gap} = \overline{V}_{gap} / d$. The voltage divider rule is applied in the receiving mode of the antenna (see inset of Figure 7(a)) in order to derive the voltage drop across the gap in the presence of an input plane wave polarized along the antenna axis with electric field amplitude $E_0 = 1 \text{ V/m}$. Hence the voltage across the gap is $\overline{V}_{gap} = \overline{V}_{OC} Z_{gap} / (Z_{gap} + Z_{dip})$, where the open circuit voltage $\overline{V}_{OC} = E_0 L_{eff}$ depends on the effective antenna length, defined as in [46].

In Figures 7 (b) and (c) we plot the field enhancement, $E_{gap} / E_0$, versus gap separation and photon energy. For comparison the classical model results are shown in Figure 7(b) and the quantum corrected results are shown in Figure 7(c). The field enhancement is red shifted and continually increases in magnitude as the gap goes to zero. The QCT results also have a red shift as the gap decreases, but the enhancement factor has a maximum at a finite gap size. The field enhancement maximum is similar in size and value with that reported by Esteban et al. [19]. In our case it occurs at a gap of 0.7 nm and the field enhancement maximum value is 420 at photon energy 2 eV. The field enhancement data for Figure 7(b) and the figures that follow are summarized in Table 2. The gap in our calculations is somewhat longer than the value of 0.5 nm found in Ref [19] and could be attributed to differences in the geometry of the dimers.

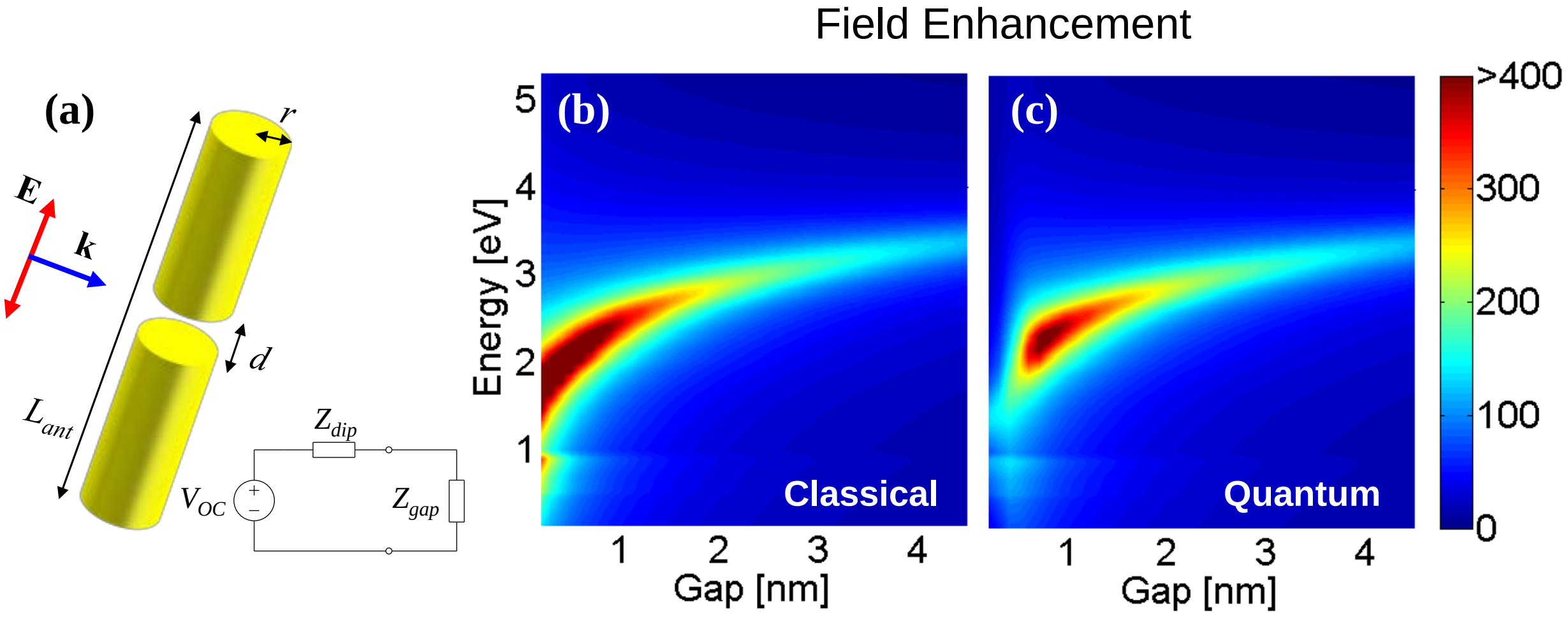

Figure 7: Field enhancement map for gold 3D cylinder dimers with a vacuum gap. (a) The cylindrical, center-fed nanoantenna geometry; (b): The classical electromagnetic model; (c) the QCT applied to the vacuum gap.

Next we consider two parallel, infinitely long, gold cylinders. A plane wave is polarized with the electric field parallel to symmetry axis passing through the centers of the two cylinders and it is incident from the side perpendicular to that axis. Linear and nonlinear conductivities in the gap region are assigned by adopting a simple methodology. We draw straight lines connecting the two gold nanoparticles along the symmetry axis, as illustrated in Figure 8(a). The conductivities on that line are assigned according to its length by identifying the length with the gap parameter; near the symmetry axis the conductivity is highest and so is the tunneling current, which is illustrated as a thick line in Fig. 8(a). The line thickness becomes thinner indicating that the current drops off as the effective gap widens. Under this assumption, the quantum ac conductivity defined in Eqs. (23) as a function of the inter-particle distance $d$ and the frequency $\omega$, can be written as $\sigma(\omega, d)$, where $d = s + 2r\left\{1-\cos\left[\sin^{-1}(x/r)\right]\right\}$ and $s$ is the minimum separation (or gap parameter) between the particles, i.e., the inter-particle distance at $x = 0$.

A 2D Comsol finite element solver is used to generate Figures 8 through 10, where the field enhancement is mapped for the gold dimer in vacuum and embedded in two different insulators. The field enhancement for the three cases shows several resonance peaks due to plasmon hybridized excitations [48]. In the hybridation scheme the symmetric resonances are strongly red shifted as the gap is decreased. The peak field enhancement of 420 occurs for the lowest energy hybrid resonance at 2 eV for a gap of 0.7 nm. The field enhancement is quenched below 0.5 nm. Our smallest gap in the figures is 0.2 nm.

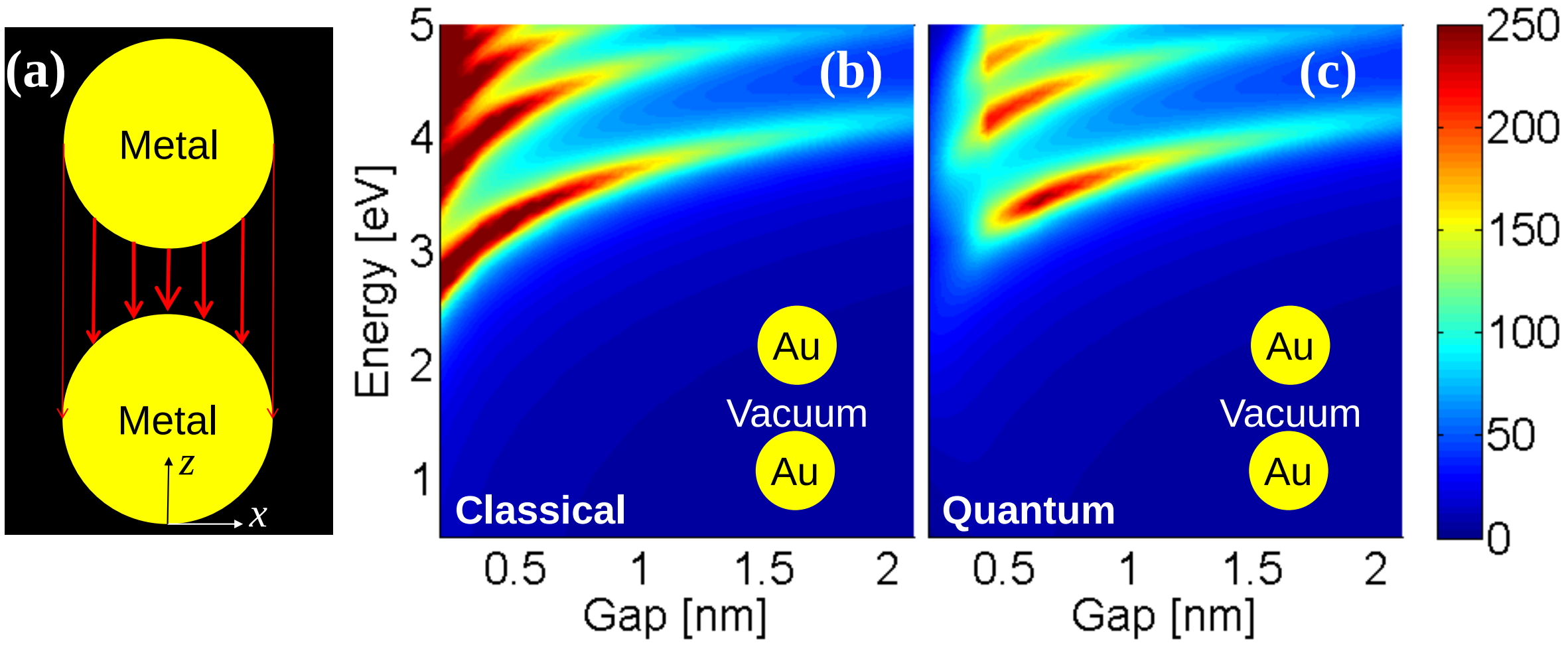

Figure 8: Enhanced field maps for gold 2D dimer cylinders embedded in vacuum. The cylinder radii are 10 nm. (a) An illustration of a simple method for assigning linear and nonlinear conductivities in the gap region. The quantum conductivity values are assigned along a straight line connecting the cylinders; and the length of the line is set to the corresponding gap parameter. The current is viewed as flowing along the lines connecting the two particles so that thicker arrows show greater current flow that tapers off at the edges due the lower conductivity. (b) Map of the field enhancement using the classical approach. (c) Field enhancement map using QCT. The minimum gap separation is 0.2 nm.

For the gold dimer in vacuum (Figure 8) the lowest energy hybrid resonance displays the same general characteristics as found with the gold nanoantenna in Figure 7. The peak field enhancement red shifted as the gap parameter is reduced and the maximum field enhancement value, 244, is found for photon energy 3.5 eV and gap parameter 0.66 nm; by comparison with the gold dimer embedded in silica, Figure 9, the resonances are shifted to lower energies due to the dielectric constant of the medium. In this case, even though the electron affinity of silica is small, the peak of the field enhancement is 263 at d=0.45 nm and photon energy 2.3 eV. This is a shorter gap parameter than found for the case of the gold dimer in vacuum. These results are summarized in Table 2.

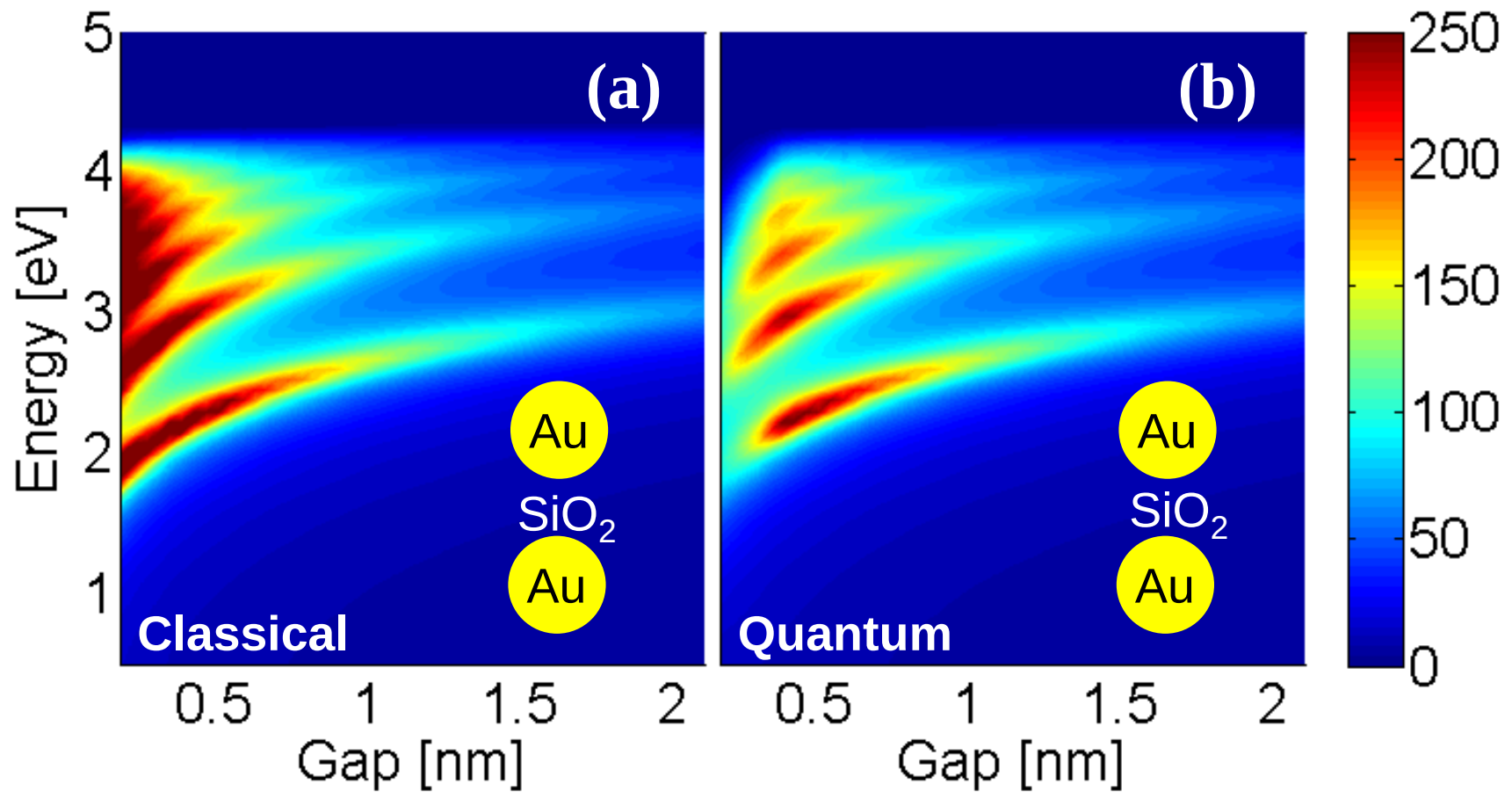

Figure 9: Enhanced field maps for gold 2D dimer cylinders embedded in silica. (a) The field enhancement using classical approach. (b) Field enhancement using QCT. The cylinder radii are 10 nm.

For the gold dimer embedded in titanium dioxide the resonance energies are the lowest of the three cases. The multiple resonance structure is retained and the field enhancement is responsible for reducing the potential barrier which increases the quantum ac conductivity. The peaks in the enhancement occur at the edge of our separation boundary of 0.2 nm. The peak value of the field enhancement is 210 at the photon energy 1.7 eV for the second hybridation resonance at a gap parameter at 0.3 nm (Table 2). The maximum field enhancement is lower than we observed for the other cases, but the value depends on the dielectric function of the gold, which is highly dispersive. The lowest energy resonances are found for the gold dimer in titanium dioxide; this is a consequence of the dielectric constant being the largest for titanium dioxide (see Table 1).

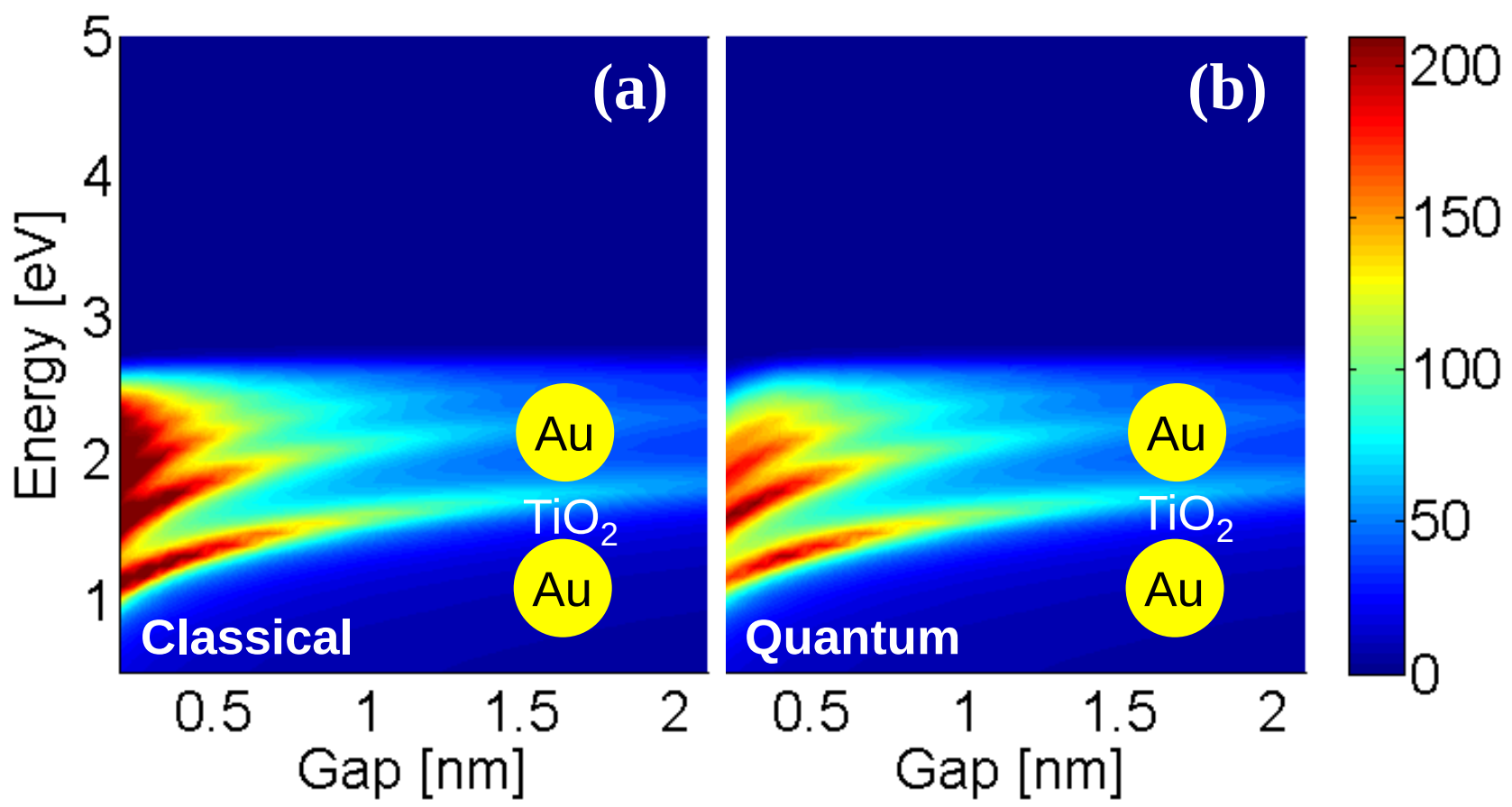

Figure 10: Enhanced field maps for gold 2D dimer cylinders embedded in titanium dioxide. (a) The field enhancement using classical approach. (b) Field enhancement using QCT. The cylinder radii are 10 nm.

### *5.3 Application to Sodium Dimers*

For the next example we apply QCT to the same two-dimensional dimer geometry but now the cylinders are made of sodium. For this study we only consider the vacuum environment. The current density and nonzero conductivities are mapped in Figure 11 as a function of the gap parameter and applied voltage ($J_{dc}$) or wavelength (conductivities). The general characteristics of these physical quantities are similar to the results for gold cylinders. However, sodium has a smaller work function which increases the corresponding dc current densities for the same gap and applied voltage. The smaller work function also produces correspondingly larger quantum conductivities, as well. Ultimately we examine its effect on the electromagnetic properties of the nanostructures.

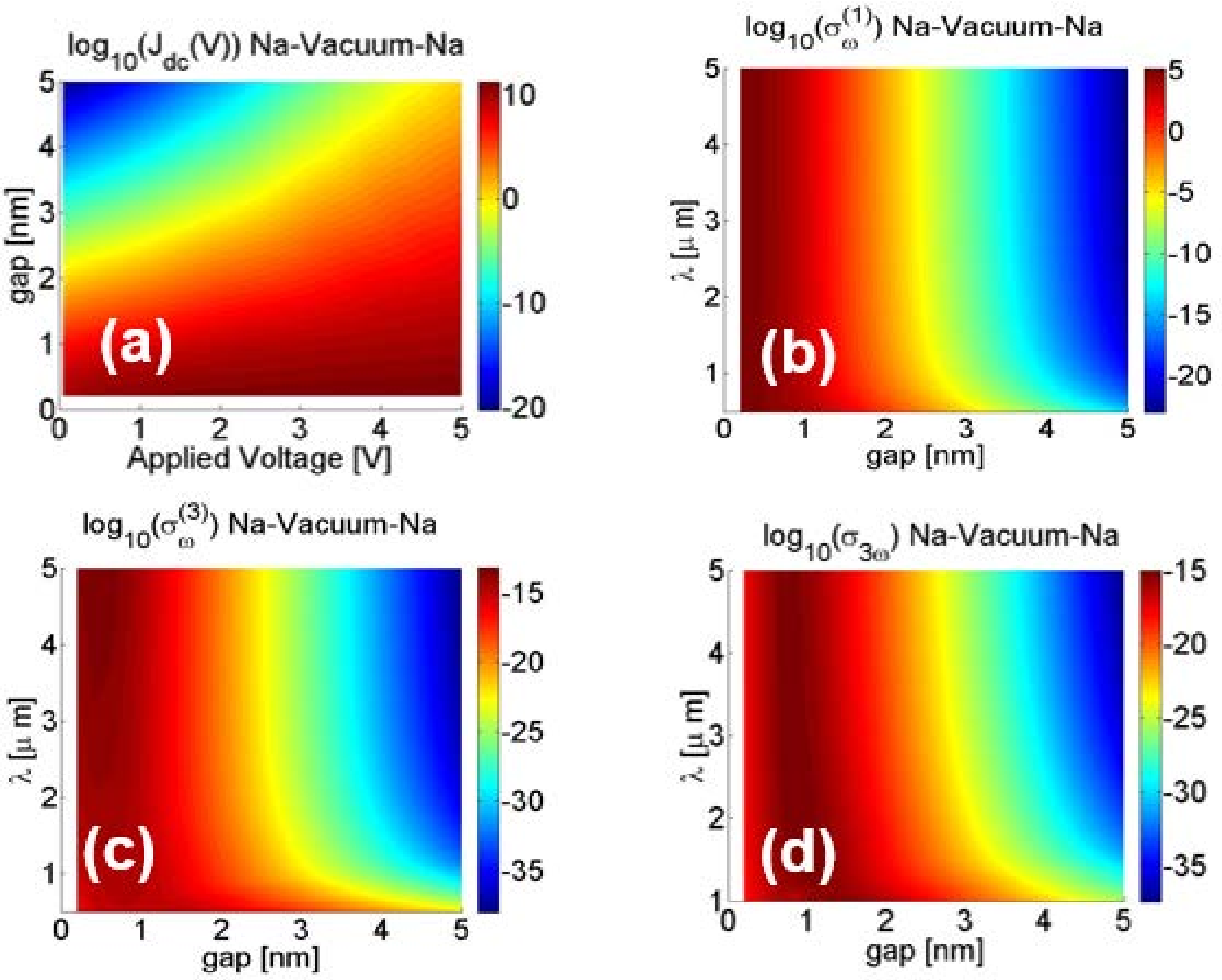

Figure 11: Maps of the logarithm (base 10) of several physical quantities for a Na-Vacuum-Na dimer. (a) The tunneling current; (b) the linear ac conductivity; (c) the nonlinear TPA conductivity; (d) the nonlinear THG conductivity.

The sodium quantum conductivities are applied to the two-dimensional system of sodium cylindrical nanowires with radius $r$ = 4.9 nm. The radius was chosen to correspond to results in Ref. [22]. A plane wave with electric field oriented along the dimer axis, as shown in the inset of Figure 12(a), impinges on the nanoparticles' system. The results for the field enhancement at the center of the gap under two different situations are presented in Figure 12. In Figure 12(a) the field enhancement in the linear regime, i.e., with low input intensities, for a range of photon energies and gap separations is reported. The peak field enhancement is 55 at the photon energy of 3.22 eV and the gap is 1.1 nm. The results reported by Teperik et al. [22] using time-dependent density functional theory are similar with a field enhancement peaks near 80 at a photon energy near 3 eV with the gap near 0.6 nm. Our gap is somewhat larger than found in their study; the difference may be due to a discrepancy in the value of the work function.

The question as to the size effect of the optical nonlinearities can be answered by incorporating nonlinearities into the Comsol simulation. To showcase an initial nonlinear result we present the effect of the TPA conductivity on the field enhancement. In Fig. 12(b) the irradiance of the incident field was taken as 1 GW/cm$^2$. The field enhancement maximum is reduced to 36 at the photon energy, 3.4 eV, and the peak position has moved to a larger gap size, 1.75 nm (see Table 2). By comparison with the map in Fig. 12(a) representing the linear result the peak was further reduced by 35% due to the quantum nonlinearity. This result is intriguing since it is comparable in magnitude to the nonlinearity effect reported in Ref. [18] for gold nanospheres. However, this topic deserves a more complete and thorough analysis will be reserved for the future.

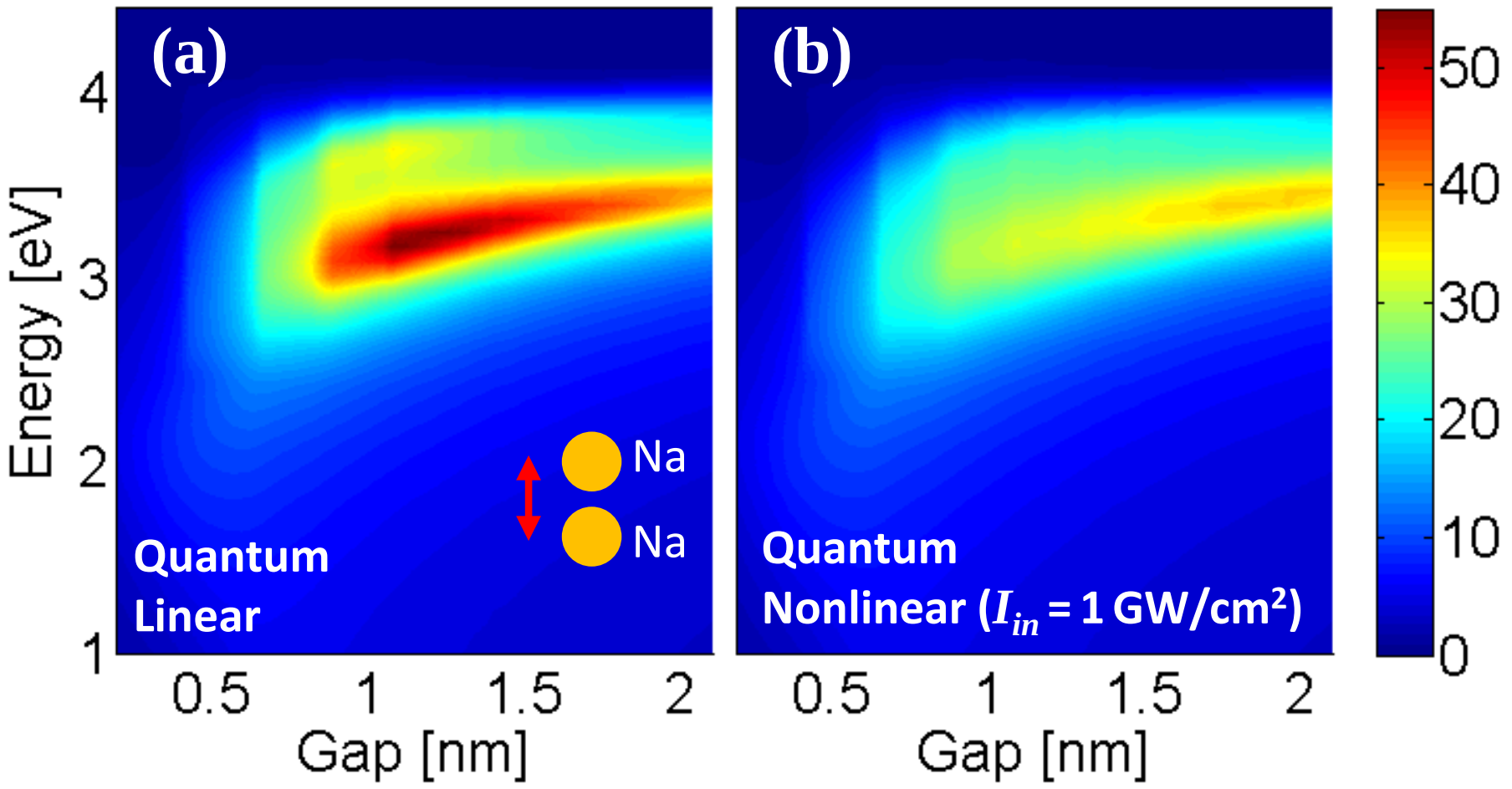

Figure 12: Enhanced field maps for sodium dimer cylinders using the quantum conductivity model. (a) The field enhancement using the linear ac quantum conductivity term. The red arrow in the inset shows the electric field polarization of the input plane wave. (b) Field enhancement adding the nonlinear conductivity term, $\sigma_{\omega}^{(3)}$. The applied field irradiance is 1 GW/cm$^2$.

Table 2: Summary of field enhancement maxima ($FE_{max}$) data extracted from the maps in the selected figures.

| Figure number | $FE_{max}$ | Gap [nm] | Energy [eV] |
|---|---|---|---|
| 7(c) | 420 | 0.7 | 2 |
| 8(c) | 244 | 0.66 | 3.5 |
| 9 | 263 | 0.45 | 2.3 |
| 10 | 210 | 0.3 | 1.7 |
| 12(a) | 55 | 1.1 | 3.22 |
| 12(b) | 36 | 1.75 | 3.4 |

## 6. CONCLUSIONS

Applying the quantum theory of tunneling for MIM structures we derived a set of linear and nonlinear quantum conductivities that can be employed to predict optical properties of nanoscale plasmonic systems. All physical parameters used in our calculations are extracted from the literature and no additional assumptions or fit parameters are required. Furthermore QCT can provide results for any combinations of metals and insulators, and will enable the design of complex device structures with optimized electromagnetic performance. The quantum conductivities also predict new, nonlinear optical effects, which could be exploited in developing future photonic devices.

Using QCT we derive three new nonlinear ac conductivities that can be detected in carefully designed experiments. The description of nonlinear coefficients can be extended to higher harmonic generation to study its efficiency for EUV production, as well. As a validation of QCT we demonstrated how the linear ac conductivity reduces the field enhancement, which compared favorably with results using different methods that were recently reported in several publications [18-22]. Our nonlinear quantum conductivity for two-photon absorption affects the electromagnetic field in the same way as the linear ac conductivity. Our TPA conductivity coefficient suppresses the field enhancement as the electromagnetic irradiance is increased and it has a similar magnitude to that reported from numerical calculations [18]. In addition we also predict a third-harmonic field can be generated from the quantum tunneling effect. The second-harmonic coefficient vanishes for dimers made with the same metals and no applied field breaks the symmetry, but by breaking this symmetry though a second-harmonic dipolar field can be generated. The third-harmonic term is nonzero for all combinations of metals, insulators and applied field. Future studies need to explore nonlinear phenomena in greater detail.

## ACKNOLEDGEMENT

This research was performed while the authors J. W. Haus, M. A. Vincenti and D. de Ceglia held a National Research Council Research Associateship award at the U.S. Army Aviation and Missile Research Development and Engineering Center. J.W. Haus also thanks Concita Sibilia and Marco Centini for stimulating discussions while visiting the University of Rome La Sapienza.

## REFERENCES

[1] H. Raether, *Surface plasmons on smooth and rough surfaces and on gratings,* volume **111** (Springer Verlag, Berlin, 1988).

[2] H. A. Atwater and A. Polman, "Plasmonics for improved photovoltaic devices," Nat Mater **9**, 205 (2010).

[3] P. Bharadwaj, B. Deutsch and L. Novotny, "Optical antennas," *Adv. Opt. Photon.* **1**, 438 (2009).

[4] S. Hayashi and T. Okamoto, "Plasmonics: visit the past to know the future," J. Phys. D 45, 433001 (2012).

[5] N. J. Halas, L. Surbhi, W.-S. Chang, S. Link and P. Nordlander, "Plasmons in Strongly Coupled Nanostructures," Chem. Rev. **111**, 3913 (2011).

[6] N. C. Nyquist, P. Nagpal, K. M. McPeak, D. J. Norris and S.-H. Oh, "Engineering metallic nanostructures for plasmonics and nanophotonics," Rep. Prog. Phys. **75**, 036501 (2012).

[7] P. Biagioni, J.-S. Huang and B. Hecht, "Nanoantennas for visible and infrared radiation," Rep. Prog. Phys. **75**, 024402 (2012).

[8] Z. Han and S. I. Bozhevolnyi, "Radiation guiding with surface plasmon polaritons," Rep. Prog. Phys. 76, 016402 (2013).

[9] J. N. Anker, W. P. Hall, O. Lyandres, N. C. Shah, J. Zhao and R. P. Van Duyne, "Biosensing with plasmonic nanosensors," *Nature Mater.* **7**, 442 (2008).

[10] M. L. Brongersma, R. Zia and J. A. Schuller, "Plasmonics – the missing link between nanoelectronics and microphotonics," J Applied Physics A **89**, 221-223 (2007).

[11] N. Aközbek, N. Mattiucci, D. de Ceglia, R. Trimm, A. Alù, G. D'Aguanno, M. Vincenti, M. Scalora and M. Bloemer, "Experimental demonstration of plasmonic Brewster angle extraordinary transmission through extreme subwavelength slit arrays in the microwave," Phys. Rev. B **85**, 205430 (2012).

[12] M. Grande, G. V. Bianco, M. A. Vincenti, T. Stomeo, D. de Ceglia, M. De Vittorio, V. Petruzzelli, M. Scalora, G. Bruno and A. D'Orazio, "Experimental surface-enhanced Raman scattering response of two-dimensional finite arrays of gold nanopatches," Appl. Phys. Lett. **101**, 111606 (2012).

[13] W. Rechberger, A. Hohenau, A. Leitner, J. R. Krenn, B. Lamprecht and F. R. Aussenegg, "Optical properties of two interacting gold nanoparticles," Opt. Comm. **220,** 137 (2003).

[14] A. Aubry, D. Y. Lei, S. A. Maier and J. B. Pendry, "Interaction between Plasmonic Nanoparticles Revisited with Transformation Optics," Phys. Rev. Lett. **105**, 2339010-4 (2010).

[15] S. Kim, J Jin, Y.-J. Kim, I.-Y. Park, Y. Kim and S.-W. Kim, "High-harmonic generation by resonant plasmon field enhancement," Nature **453**, 757 (2008).

[16] J. Zuloaga, E. Prodan and P. Nordlander, "Quantum description of the plasmon resonances of a nanoparticle dimer," Nano Lett. **9,** 887 (2009).

[17] J. Zuloaga, E. Prodan and P. Nordlander, "Quantum plasmonics: Optical properties and tunability of metallic nanorods," ACS Nano **4**, 5269 (2010).

[18] D.C. Marinica, A.K. Kazansky, P. Nordlander, J. Aizpurua and A.G. Borisov, "Quantum plasmonics: nonlinear effects in the field enhancement of a plasmonic nanoparticle dimer." Nano Lett. **12**, 1333 (2012).
[19] R. Esteban, A.G. Borisov, P. Nordlander and J. Aizpurua, "Bridging quantum and classical plasmonics with a quantum-corrected model." Nature Comm. **3**, 825 (2012).
[20] R. Alvarez-Puebla, L. M. Liz-Marzan and F. J. Garcia de Abajo, "Light concentration at the nanometer scale, J. Phys. Chem. Lett. **1**, 2428 (2010).
[21] C. Ciraci, R. T. Hill, J. J. Mock, Y. Urzhumov, A. I. Fernandez-Dominguez, S. A. Maier, J. B. Pendry, A. Chilkoti and D. R. Smith, "Probing the ultimate limits of plasmonic enhancement," Science **337**, 1072 (2012).
[22] T.V. Teperik, P. Nordlander, J. Aizpurua and A. G. Borisov, "Quantum Plasmonics: Nonlocal effects in coupled nanowire dimer," arXiv:1302.3339 [physics.optics].
[23] C.Fumeaux, W. Herrmann, F. K. Kneubühl and H. Rothuizen, "Nanometer thin-film Ni–NiO–Ni diodes for detection and mixing of 30 THz radiation," Infrared Phys. Technol. **39**, 123 (1998).
[24] M. R. Abdel-Rahman, F. J. Gonzalez, G. Zummo, C. F. Middleton and G. D. Boreman, "Antenna-coupled MOM diodes for dual-band detection in MMW and LWIR," Proc. SPIE **5410**, 233 (2004).
[25] P. C. Hobbs, R. B. Laibowitz, F. R. Libsch, N. C. LaBianca and P. P. Chiniwalla, "Efficient waveguide-integrated tunnel junction detectors at 1.6 μm," Opt Express **15**,16376 (2007).
[26] M. Nagae, "Response time of metal-insulator-metal tunnel junctions," Jpn. J. Appl. Phys. **11**, 1611 (1972).
[27] W. Tantraporn, "Electron current through metal-insulator-metal sandwiches," Solid-State Electronics **7**, 81 (1964).
[28] L. O. Hocker, D. R. Sokoloff, V. Daneu and A. Javan, "Frequency mixing in the infrared and far-infrared using a metal-to-metal point contact diode," Appl. Phys., **12**, 401 (1968).
[29] A. Sanchez, C. F. Davis, K. C. Liu and A. Javan, "The MOM tunneling diode: theoretical estimate of its performance at microwave and infrared frequencies," J. Appl. Phys. **49,** 5270 (1978).
[30] C. Fumeaux, W. Herrmann, F. K. Kneubühl and H. Rothuizen. "Nanometer thin-film Ni–NiO–Ni diodes for detection and mixing of 30 THz radiation," Infrared Phys. Tech. **39**, 123 (1998).
[31] M. Dagenais, K. Choi, F. Yesilkoy, A. N. Chryssis and M. C. Peckerar, "Solar spectrum rectification using nano-antennas and tunneling diodes," Proc. SPIE **7605**, 76050E (2010).
[32] S. Bhansali, S. Krishnan, E. Stefanakos and D. Y. Goswami, "Tunneling junction based rectenna - a key to ultrahigh efficiency solar/thermal energy conversion," AIP Conf. Proc. **1313**, 79 (2010).
[33] S. Grover, O. Dmitriyeva, M. J. Estes and G. Moddel, "Traveling-wave metal/insulator/metal diodes for improved infrared bandwidth and efficiency of antenna coupled rectifiers," Nanotechnol IEEE Trans. **9**,716 (2010).

[34] S. Grover and G. Moddel, "Engineering the current–voltage characteristics of metal–insulator–metal diodes using double-insulator tunnel barriers," Solid-State Electronics **67**, 94 (2012).
[35] S. Grover and G. Moddel, "Applicability of metal/insulator/metal (MIM) diodes to solar rectennas," IEEE J. of Photovoltaics **1**,78 (2011).
[36] J. W. Haus, L. Li, N. Katte, C. Deng, M. Scalora, D. de Ceglia and M. A. Vincenti, "Nanowire Metal-Insulator-Metal Plasmonic Devices," ICPS 2013: International Conference on Photonics Solutions, edited by P. Buranasiri, S. Sumriddetchkajorn, Proc. of SPIE **8883**, 888303 (2013).
[37] H. Kroemer, *Quantum Mechanics*, 3rd ed. (Prentice-Hall Inc., NJ, 1994).
[38] J. G. Simmons, "Generalized Formula for the Electric Tunnel Effect between Similar Electrodes Separated by a Thin Insulating Film," J. Appl. Phys. **34**, 1793 (1963).
[39] J. G. Simmons, "Electric Tunnel Effect between Dissimilar Electrodes Separated by a Thin Insulating Film," J. Appl. Phys. **34**, 2581 (1963).
[40] R. J. Whitefield and J. J. Brady, "New Value for Work Function of Sodium and the Observation of Surface-Plasmon Effects," Phys. Rev. Lett. **26**, 380 (1971). Erratum: Phys. Rev. Lett. **26**, 1005 (1971).
[41] J. Robertson, "Band offsets of high dielectric constant gate oxides on silicon," J. Non-Crystalline Sol. **303**, 94 (2002).
[42] P. K. Tien and J. P. Gordon, "Multiphoton Process Observed in the Interaction of Microwave Fields with the Tunneling Between Superconductor Films," Phys. Rev. **129**, 647 (1963).
[43] J. R. Tucker and M. F. Millea, "Photon detection in nonlinear tunneling devices," Appl. Phys. Lett. **33**, 611 (1978).
[44] J. R. Tucker and M. J. Feldman, "Quantum detection at millimeter wavelengths," Rev. Mod. Phys. **57**, 1055 (1985).
[45] J. R. Tucker , "Quantum Limited Detection in Tunnel Junction Mixers," IEEE J. of Quant. Electron. **QE-15**, 1234 (1979).
[46] A. Locatelli, C. De Angelis, D. Modotto, S. Boscolo, F. Sacchetto, M. Midrio, A.-D. Capobianco, F. M. Pigozzo, and C. G. Someda, "Modeling of enhanced field confinement and scattering by optical wire antennas," Opt. Express **17**, 16792 (2009).
[47] C. A. Balanis, *Antenna theory: analysis and design* (Wiley, NY, 2005).
[48] P. Nordlander, C. Oubre, E. Prodan, K. Li and M. I. Stockman, Nano Lett. **4**, 899 (2004).